\documentclass{PoS}
\usepackage{epsfig}
\def\be{\begin{equation}}
\def\ee{\end{equation}}
\def\q{q{\bar q}}
\def\Q{Q{\bar Q}}
\def\tx{\tilde x}
\def\tu{\tilde t}
\def\lsim{\raise0.3ex\hbox{$<$\kern-0.75em\raise-1.1ex\hbox{$\sim$}}}
\def\NP{{ Nucl.\ Phys.\ }}
\def\PL{{ Phys.\ Lett.\ }}
\def\PR{{ Phys.\ Rev.\ }}
\def\PRL{{ Phys.\ Rev.\ Lett.\ }}
\def\ZP{{ Z.\ Phys.\ }}

\title{Thermal Hadron Production by QCD Hawking Radiation}

\ShortTitle{Thermalization in strong interactions}

\author{{Helmut Satz}\\
~\\
         University of Bielefeld, Germany\\
         E-mail: \email{satz@physik.uni-bielefeld.de}}

\abstract{The QCD counterpart of Hawking radiation from black holes leads
to thermal hadron production in high energy collisions, from $e^+e^-$ 
annihilation to heavy ion interactions. This hadronic radiation 
is formed by tunnelling through the event horizon of colour confinement
and is emitted at a universal temperature $T_H \simeq (\sigma /2 \pi)^{1/2}$,
where  $\sigma$ denotes the string tension. Since the event horizon
does not allow information transfer, the radiation is thermal
``at birth''.
}

\FullConference{Critical Point and Onset of Deconfinement\\
                 July 3-6 2006\\
                 Florence, Italy}

\begin{document}

\section{Introduction}
Quantum chromodynamics, through colour confinement, restricts quarks and 
gluons to move in a limited region of space. In a black hole, matter and 
light are confined by gravitation to remain within a restricted region of 
space. This similarity was noted quite soon after the advent of QCD, and 
it was suggested that hadrons were the analogue of black holes in strong 
interaction physics \cite{C-R,Salam,Grillo}. 
In the case of gravitational black holes, the 
isolation of the system is not quite absolute: in the strong field at the 
outer edge, quantum excitation can lead to the emission of Hawking radiation 
into the physical vacuum \cite{Hawking}. Since no information transfer 
between the inside and the outside of the black hole is allowed, 
this radiation must give equal {\sl a priori} weights to all possible 
states on the outside, and hence it is thermal at the point of formation.   
In this report, based on joint work with Paolo Castorina and Dmitri 
Kharzeev \cite{CKS}, we want to show that 
high energy collisions in strong interaction physics produce a self-similar 
cascade of ``white holes'' (colourless from the outside, but coloured 
inside), and that the Hawking radiation arising at the confinement horizon 
of these can provide the thermal behaviour observed quite universally in all 
soft hadron production \cite{Hagedorn,species}.

We begin by recalling briefly the relevant basic aspects of black hole 
physics and indicate some first relations to strong interactions. We then 
turn to high energy $e^+e^-$ annihilation as the simplest hadron production 
process and study the effect of quark-antiquark pair excitation and string 
breaking. Next we review some features of charged and of rotating  
black holes; these allow us to generalize our scenario to hadronic 
collisions at finite baryon density as well as to non-central interactions. 
Finally, we comment on stochastic vs.\ kinetic thermalization. 

\section{Black Holes and Hawking Radiation}

A black hole is the final stage of a neutron star after gravitational
collapse \cite{Ruf}. It has a mass $M$ concentrated in such a small volume 
that the resulting gravitational field confines all matter and even photons 
to remain inside the event horizon $R$ of the system: no causal connection 
with the outside is possible. As a consequence, black holes have three
(and only three) observable properties: mass $M$, charge $Q$ and angular 
momentum $J$. This section will address mainly black holes with $Q=J=0$;
we shall come back to the more general properties in section 4.  

Classically, a black hole would persist forever and remain forever
invisible. On a quantum level, however, its constituents (photons,
leptons and hadrons) have a non-vanishing chance to escape by tunnelling 
through the barrier presented by the event horizon. The resulting Hawking 
radiation \cite{Hawking} cannot convey any information about the internal 
state of the black hole; it must be therefore be thermal, and it was 
shown that for a non-rotating black hole of vanishing charge (denoted as 
Schwarzschild black hole), the corresponding radiation temperature is 
\be
T_{BH} = {1 \over 8 \pi~\! G~\! M}
\label{T-BH}
\ee
where $G$ is the gravitational constant. This temperature is inversely 
proportional to the mass of the black hole, and since the radiation 
reduces the mass, the radiation temperature will increase with time.
For black holes of stellar size, however, one finds $T_{BH}~\lsim~2 
\times 10^{-8}$ $^{\circ}$K. This is many orders of magnitude below the 
2.7 $^{\circ}$K cosmic microwave background, and hence not detectable. 

In general relativity, the event horizon arises as consequence of the
Schwarzschild metric and its generalizations to  $Q\not= 0, J\not= 0$.
The occurrence and role of the event horizon for thermal radiation was 
subsequently generalized by Unruh \cite{Unruh}. A system undergoing uniform 
acceleration $a$ relative to a stationary observer eventually reaches a 
classical turning point and thus encounters an event horizon. Let us recall 
the resulting hyperbolic motion \cite{Pauli}. A point mass $m$ subject to a 
constant force $F$ satisfies the equation of motion
\be
{d \over dt}{mv \over \sqrt{1-v^2}} = F,
\label{e-motion} 
\ee
where $v(t)=dx/dt$ is the velocity, normalized to the speed of light $c=1$. 
This equation is solved by the parametric form  
\be
x = {1\over a} \cosh a\tau ~~~~~ t= {1\over a} \sinh a \tau,
\label{rindler}
\ee
where $a=F/m$ denotes the acceleration in the instantaneous rest frame of 
$m$, and $\tau$ the proper time, with $d\tau=\sqrt{1-v^2} dt$. 

\begin{figure}[h]
\centerline{\psfig{file=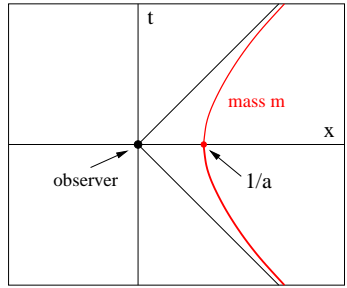,width=4.5cm}} 
\caption{Hyperbolic motion}
\label{W-L}
\end{figure}

The resulting
world line is shown in Fig.\ \ref{W-L}. It corresponds to the mass $m$
coming from $x=\infty$ at $t=-\infty$ at with a velocity arbitrarily close
to that of light, decelerating uniformly until it comes to rest at the
classsical turning point $x_H=-(1/a)$. It then accelerates again and
returns to $x=\infty$ at $t=\infty$, approaching the speed of light. 
For given $a$, $x_H$ is thus the event horizon beyond which $m$ classically 
cannot pass. The only signal an observer can detect of the passage of $m$ 
is thermal quantum radiation of temperature 
\cite{Hawking,Unruh,TD,P-W} 
\be
T_U = {a \over 2 \pi}.
\label{T-U}
\ee
In the case of gravity, we have the force
\be
F= m~\!a = G {m~\! M \over R^2},
\label{FG}
\ee
on a probe of mass $m$. From the dependence of the gravitational metric
on black hole mass and radius, we obtain
\be
R_g = 2~G~\!M
\ee
for the Schwarzschild radius of the black hole, leading to $a=1/(4~G~\! M)$ 
for the acceleration at the event horizon and thus back to eq.\ (\ref{T-BH}). 
We also see why the temperature of Hawking radiation {\sl decreases} with
increasing black hole mass: since $R_g$ grows linearly with $M$, increasing
the black hole mass decreases its energy density and hence the Hawking
temperature $T_{BH}$. 

It is instructive to consider the Schwarzschild radius of a typical hadron, 
assuming a mass $m \sim 1$ GeV:
\be
R_g^{had} \simeq 1.3 \times 10^{-38}~{\rm GeV}^{-1} \simeq 2.7 \times
10^{-39}~{\rm fm}.
\ee
To become a gravitational black hole, the mass of the hadron would thus
have to be compressed into a volume more than $10^{100}$ times smaller
than its actual volume, with a radius of about 1 fm. On the other hand,
if instead we increase the interaction strength from gravitation to strong
interaction \cite{C-R}, we gain in the resulting ``strong'' Schwarzschild 
radius $R_s^{had}$ a factor
\be
{\alpha_s \over Gm^2},
\ee
where $\alpha_s$ is the dimensionless strong coupling and $Gm^2$ the
corresponding dimensionless gravitational coupling for the case in question. 
This leads to
\be
R_s^{had} \simeq {2 \alpha_s \over m}  
\ee
which for $\alpha_s \simeq 2.5$ gives $R_s^{had} = 1$ fm. In other words,
the confinement radius of a hadron is about the size of its ``strong''
Schwarzschild radius, so that we could consider quark confinement as the 
strong interaction version of the gravitational confinement in black
holes \cite{C-R,Salam}.

The event horizon of a black hole arises from the deformation of the
Minkowski metric due to the strong gravitational force. Such a 
geometrization of force effects has also been employed in non-linear
electrodynamics, where it can lead to similar spatial 
constraints \cite{Novello}. A corresponding extension to QCD will 
be discussed in \cite{CKS}; we shall
here concentrate on the more phenomenological aspects.  

\section{Pair Production and String Breaking}

In the previous section, we had considered constituents of a black 
hole undergoing accelerated motion in classical space-time.
In this section, we shall first address the modifications which arise if 
the underlying space-time manifold is specified by quantum field theory,
so that in the presence of a strong field the vacuum becomes unstable 
under pair production. Next we turn to the specific additional features 
which come in when the basic constituents are subject to colour confinement
and can only exist in colour neutral bound states. 

\begin{figure}[h]
\centerline{\psfig{file=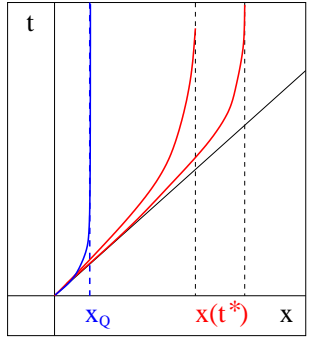,width=4cm}} 
\caption{Classical and quantum horizons in $\q$ separation}
\label{TP}
\end{figure}

As basic starting point, we consider two-jet $e^+e^-$ annihilation 
at cms energy $\sqrt s$, 
\be
e^+e^- \to \gamma \to \q \to~{\rm hadrons}.
\label{two-jet}
\ee  
The initially produced $\q$ pair flies apart, subject to the constant 
confining force given by the string tension $\sigma$; this  
results in hyperbolic motion \cite{Barshay,Hosoya,Horibe} of the type 
discussed 
in the previous section. At $t=0$, the $q$ and $\bar q$ separate with an 
initial velocity $v_0 = p/\sqrt{p^2 + m^2}$, where $p \simeq \sqrt s/2$ 
is the momentum of the primary constituents in the overall cms and $m$ 
the effective quark mass. We now have to solve Eq.\ (\ref{e-motion}) 
with this situation as boundary condition; the force
\be
F = \sigma,
\label{sigma}
\ee
is given by the string tension $\sigma$ binding the $\q$ system. The
solution is
\be
\tx = [1 - \sqrt{1-v_0\tu + \tu^2}]
\label{sol1}
\ee
with $\tx=x/x_0$ and $\tu=t/x_0$; here the scale factor 
\be
x_0 = {m \over \sigma} {1\over \sqrt{1-v_0^2}} = {1 \over a}~\gamma
\label{sol2}
\ee
is the inverse of the acceleration $a$ measured in the overall cms. The 
velocity becomes
\be
v(t) = {dx \over dt} = {(v_0/2) - \tu \over \sqrt{1 - v_0\tu + \tu^2}};
\label{sol3}
\ee
it vanishes for
\be
\tu^* = {v_0\over 2} ~\Rightarrow~ t^* = {v_0\over 2}~{m\over \sigma}~\gamma,
\label{sol4}
\ee
thus defining
\be
x(t^*) = {m\over \sigma}~\gamma~\left( 1 - \sqrt{1-(v_0^2/4)}\right) 
\simeq {\sqrt s \over 2 \sigma}
\label{sol5}
\ee
as classical turning point and hence as the classical event horizon measured 
in the overall cms (see Fig. \ref{TP}). 

Eq.\ (\ref{sol5}) allows the $q$ and the $\bar q$ to separate arbitrarily
far, provided the pair has enough initial energy; this clearly violates 
colour confinement. Our mistake was to consider the $\q$ system
in a classical vacuum; in quantum field theory, the vacuum itself contains
virtual $\q$ pairs, and hence it is not possible to increase the potential 
energy of a given $\q$ state beyond the threshold value necessary to bring 
such a $\q$ pair on-shell. In QED, the corresponding phenomenon was addressed
by Schwinger \cite{Schwinger}, who showed that in the presence of a constant
electric field of strength $\cal E$ the probability of producing an
electron-positron pair is given by
\be
P(M,{\cal E}) \sim \exp\{-\pi m_e^2 / e {\cal E}\},
\label{schwinger}
\ee
with $m_e$ denoting the electron mass and $e$ denoting the electric charge. 
In QCD, we expect a similar effect when the string tension exceeds the pair 
production limit, i.e., when
\be
\sigma~x > 2~\!m
\label{pair}
\ee
where $m$ specifies the effective quark mass. Beyond this point, any further 
stretching of the string is expected to produce a $\q$ pair with the 
probability
\be
P(M,\sigma) \sim \exp\{-\pi m^2 / \sigma\},
\label{string-pair}
\ee
with the string tension $\sigma$ replacing the electric field strength
$e\cal E$. This string breaking acts like a quantum event horizon 
$x_q = 2~\!m / \sigma$, which becomes operative long before the classical 
turning point is ever reached (see Fig.\ \ref{TP}). Moreover, the resulting 
allowed separation distance for our $\q$ pair, the colour confinement 
radius $x_Q$, does not depend on the initial energy of the primary quarks. 

There are some important differences between QCD and QED. In case of the 
latter, the initial electric charges which lead to the field $\cal E$ can 
exist independently in the physical vacuum, and the produced pair can be 
simply ionized into an $e^+$ and an $e^-$. In contrast, neither the primary
quark nor the constituents of the $\q$ pair have an independent existence, 
so that in string breaking colour neutrality must be preserved. As a result,
the Hawking radiation in QCD must consist of $\q$ pairs, and these can
be produced in an infinite number of different excitation states of 
increasing mass and degeneracy. Moreover, the $\q$ pair spectrum is
itself determined by the strength $\sigma$ of the field, in contrast
to the exponent $m_e^2/\cal E$ in eq.\ (\ref{schwinger}), where the value
of $\cal E$ has no relation to the electron mass $m_e$.  
 
Hadron production in $e^+e^-$ annihilation is believed to proceed in 
the form of a self-similar cascade \cite{bj,nus}. Initially, we have the
separating primary $\q$ pair, 
\be
\gamma \to [\q]
\label{bj1}
\ee
where the square brackets indicate colour neutrality. When the energy of
the resulting color flux tube becomes large enough, a further pair 
$q_1\bar q_1$ is excited from the vacuum, 
\be
\gamma \to [q [\bar q_1 q_1] \bar q].
\label{bj2}
\ee
Although the new pair is at rest in the overall cms, each of its 
constituents has a transverse momentum $k_T$ determined, through 
the uncertainty relation, by the transverse dimension $r_T$ of the flux 
tube. The slow $\bar q_1$ now screens the fast primary $q$ from its 
original partner $\bar q$, with an analoguous effect for the $q_1$ and
the primary antiquark. To estimate the $\q$ separation distance at the
point of pair production, we recall that the basic thickness of the 
flux tube connecting the $\q$ pair is in string theory given by 
\cite{Luescher}
\be
r_T = \sqrt{2\over \pi \sigma};
\label{thick}
\ee
higher excitations lead to a greater thickness and eventually to a 
divergence (the ``roughening'' transition). From the uncertainty relation 
we then have
\be
k_T \simeq \sqrt{\pi \sigma \over 2}.
\label{kt}
\ee 
With this transverse energy is included in eq.\ (\ref{pair}), we obtain 
for the pair production separation $x_Q$ 
\be
\sigma x_q = 2\sqrt{m^2 + k_T^2} ~~\Rightarrow~~x_q \simeq 
{2~\over\sigma}\sqrt{m^2 + (\pi \sigma/2)} \simeq \sqrt{2\pi \over \sigma}
\simeq 1.1~{\rm fm},
\label{x-H}   
\ee
with $\sigma = 0.2$ GeV$^2$ and $m^2 \ll \sigma$. 

\begin{figure}[h]
\centerline{\psfig{file=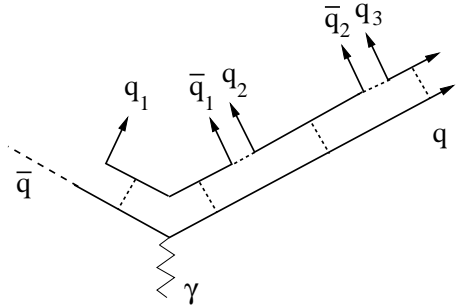,width=6cm}} 
\caption{String breaking through $\q$ pair production}
\label{breaking}
\end{figure}

Once the new pair is present, we have a colour-neutral system 
$q\bar q_1 q_1 \bar q$; but since there is a sequence of connecting
string potentials $q \bar q_1$, $\bar q_1 q_1$ and $q_1 \bar q$,
the primary string is not yet broken. To achieve that, the binding
of the new pair has to be overcome, i.e., the $q_1$ has to tunnel through
the barrier of the confining potential provided by $\bar q_1$, and 
vice versa. Now the $q$ excerts a longitudinal force on the $\bar q_1$, 
the $\bar q$ on the $q_1$, resulting in a longitudinal acceleration and 
ordering of $q_1$ and $\bar q_1$. When (see Fig.\ \ref{breaking})
\be
\sigma x(q_1\bar q_1) = 2\sqrt{m^2 + k_T^2},
\ee
the $\bar q_1$ reaches its $q_1\bar q_1$ horizon; on the other hand, when
\be
\sigma x(q \bar q_1) = 2\sqrt{m^2 + k_T^2},
\ee
the new flux tube $q \bar q_1$ reaches the energy needed to produce a
further pair $q_2 \bar q_2$. The $\bar q_2$ screens the primary $q$
from the $q_1$ and forms a new flux tube $q \bar q_2$. At this point,
the original string is broken, and the remaining pair $\bar q_1 q_2$ 
form a colour neutral bound state which is emitted as Hawking radiation
in the form of hadrons, with the relative weights of the different states
governed by the corresponding Unruh temperature. The resulting pattern is 
schematically illustrated in Fig.\ \ref{breaking}.

To determine the temperature of the hadronic Hawking radiation, we 
return to the original pair excitation process. To produce a quark of
momentum $k_T$, we have to bring it on-shell and change its velocity
from zero to $v = k_T / (m^2 + k_T^2)^{1/2} \simeq 1$. This has to
be achieved in the time of the fluctuation determined by the virtuality
of the pair, $\Delta \tau = 1/\Delta E \simeq 1/ 2k_T$. The resulting
acceleration thus becomes
\be
a = {\Delta v \over \Delta \tau} \simeq 2~\!k_T \simeq \sqrt{2 \pi \sigma}
\simeq 1.1~{\rm GeV},
\ee
which leads to 
\be
T_H = {a\over 2 \pi} \simeq \sqrt{\sigma \over 2 \pi} \simeq 180~{\rm MeV}
\label{T-H}
\ee
for the hadronic Unruh temperature. It governs the momentum distribution
and the relative species abundances of the emitted hadrons.

\begin{figure}[h]
\centerline{\psfig{file=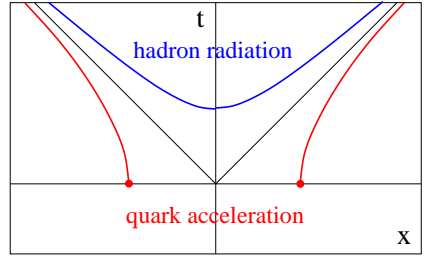,width=5.5cm}} 
\caption{Quark acceleration and hadronization world lines}
\label{world-a}
\end{figure}

A given step in the evolution of the hadronization cascade of a primary
quark or antiquark produced in $e^+e^-$ annihilation thus involves several
distinct phenomena. The color field created by the separating $q$ and 
$\bar q$ produces a further pair $q_1\bar q_1$ and then provides an
acceleration of the $q_1$, increasing its longitudinal momentum. When
it reaches the $q_1\bar q_1$ confinement horizon, still another pair
$q_2\bar q_2$ is excited; the state $\bar q_1 q_2$ is emitted as a hadron,
the $\bar q_2$ forms together with the primary $q$ a new flux tube. This
pattern thus step by step increases the longitudinal momentum of the
``accompanying'' $\bar q_i$ as well as of the emitted hadron. This, 
together with the energy of the produced pairs, causes a corresponding
deceleration of the primary quarks $q$ and $\bar q$, in order to maintain
overall energy conservation. In Fig.\ \ref{world-a}, we show the world
lines given by the acceleration $\bar q_i \to \bar q_{i+1}$ 
($q_i \to q_{i+1}$) and that of formation threshold of the hadrons 
$\bar q_i q_{i+1}$ and the corresponding antiparticles. 

\begin{figure}[h]
\centerline{\psfig{file=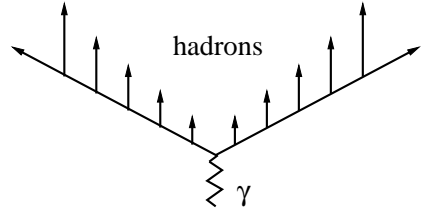,width=5.5cm}} 
\caption{Hadronization in $e^+e^-$ annihilation}
\label{multi}
\end{figure}

The energy loss and deceleration of the primary quark $q$ in this self-similar 
cascade, together with the acceleration of the accompanying partner $\bar q_i$
from the successive pairs brings $q$ and $\bar q_i$ closer and closer to 
each other in momentum, from an initial separation $q \bar q_1$ of 
$\sqrt s/2$, until they finally are combined into a hadron and the 
cascade is ended. The resulting pattern is shown in Fig.\ \ref{multi}.

The number of emitted hadrons, the multiplicity $\nu(s)$, follows quite
naturally from the picture presented here. The classical string length,
in the absence of quantum pair formation, is given by the classical
turning point determined in eq.\ (\ref{sol5}). The thickness of a flux
tube of such an ``overstretched'' string is known \cite{Luescher}; instead
of eq.\ (\ref{thick}) we have
\be
R_T^2 = {2 \over \pi \sigma} \sum_{k=0}^K {1 \over 2k + 1}
\simeq {2 \over \pi \sigma} \ln 2K,
\ee
where $K$ is the string length in units of an intrinsic string vibration
measure. From eq.\ (\ref{sol5}) we thus get
\be
R_T^2 = \simeq {2 \over \pi \sigma} \ln \sqrt s
\ee
for the flux tube thickness in the case of the classical string length. 
Because of pair production, this string breaks whenever it is stretched 
to the length $x_q$ given in eq.\ (\ref{x-H}); its thickness $r_T$ at this 
point is given by eq.\ (\ref{thick}). The multiplicity can thus be 
estimated by the ratio of the corresponding classical to quantum transverse 
flux tube areas,
\be
\nu(s) \sim {R_T^2 \over r_T^2} \simeq \ln \sqrt s,
\ee 
and is found to grow logarithmically with the $e^+e^-$ annihilation energy,
as is in fact observed experimentally.

\begin{figure}[h]
\centerline{\psfig{file=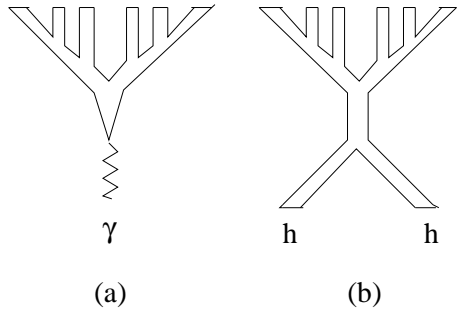,width=6cm}} 
\caption{``White hole'' structure in $e^+e^-$ annihilation (a) and 
hadronic collisions (b)}
\label{wh}
\end{figure}

\begin{figure}[h]
\centerline{\psfig{file=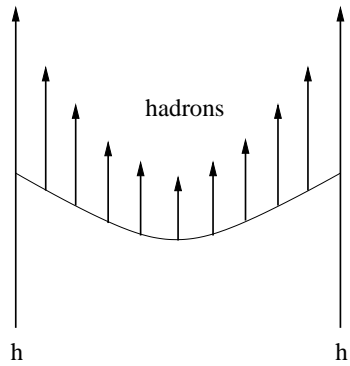,width=4.5cm}} 
\caption{Hadronization in hadron-hadron
collisions}
\label{hadromulti}
\end{figure}

Up to now, we have considered hadron production in $e^+e^-$ annihilation,
in which the virtual photon produces a confined coloured $\q$ pair as a
``white hole''. Turning now to hadron-hadron collisions, we note that here
two incident white holes combine to form a new system of the same kind,
as schematically illustrated in Fig.\ \ref{wh}. Again the resulting string 
or strong colour field produces a sequence of $\q$ pairs of increasing cms 
momentum, leading to the well-known multiperipheral hadroproduction 
cascade shown in Fig.\ \ref{hadromulti}. For the further extension to
nuclear collisions in the regime of parton saturation, see 
\cite{K-T,KLT,DdD}.

\vskip0.2cm

\medskip

In the case of hadronic, and even more so for heavy ion interactions, two 
new elements enter. The resulting system can now have an overall baryon
number, from B=2 in $pp$ to $B=400$ or more in heavy ion collisions. To
take that into account, we need to consider the counterpart of charged
black holes. Furthermore, in heavy ion collisions the resulting hadron 
production can be studied as function of centrality, and peripheral
collisions will lead to an interaction region with an overall angular
momentum. Hence we also need to consider rotating black holes. In the
next section, we therefore summarize the relevant feature of black
holes with $Q\not= 0,~J\not=0$.

\section{Charged and Rotating Black Holes}

As mentioned, for an outside observer the only characteristics of a black
hole are its mass $M$, its charge $Q$, and its angular
momentum $J$. Hence any further observables, such as the event horizon
or the Hawking temperature, must be expressable in terms of these three
quantities.

\begin{figure}[h]
\centerline{\psfig{file=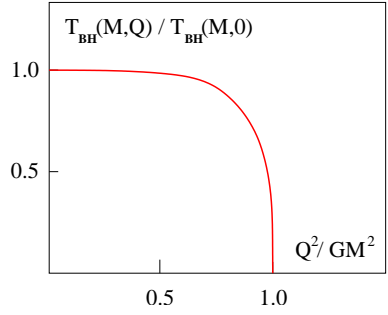,width=5cm}} 
\caption{Radiation temperature for a charged black hole}
\label{R-N}
\end{figure}

We had noted that the event horizon of a black hole is due to the strong 
gravitational attraction, which leads to a diverging Schwarzschild metric 
at a certain 
value of the spatial extension $R$. Specifically, the invariant space-time 
length element $ds^2$ is at the equator given by
\be
ds^2 = ( 1 - 2GM/R)~\!dt^2 - {1\over 1- 2GM/R}~\!dr^2,
\label{Schwarz}
\ee
with $r$ and $t$ for flat space and time coordinates; it is seen to diverge at
the Schwarzschild radius $ R_S = 2GM$. If the black hole has a net electric 
charge $Q$, the resulting Coulomb repulsion will oppose and hence weaken 
the gravitational attraction. As a result, the corresponding form
(denoted as Reissner-Nordstr\"om metric) becomes
\be
ds^2 = (1 - 2GM/R + GQ/R^2)~\!dt^2 - {1\over 1- 2GM/R + GQ/R^2}~\!dr^2.
\label{RN}
\ee
For this, the divergence leads to the smaller Reissner-Nordstr\"om radius
\be
R_{RN} = GM~\!(1 + \sqrt{1 - Q^2/GM^2}),
\label{RNradius}
\ee
which reduces to the Schwarzschild radius $R_S$ for $Q=0$. 
The temperature of the Hawking radiation now becomes \cite{Ruf,I-U-W}   
\be
T_{BH}(M,Q) = T_{BH}(M,0)~\!\left\{{2~\sqrt{1 - Q^2/ GM^2} 
\over (1 + \sqrt{1- Q^2/GM^2}~\!)^{~\!2}}\right\}; 
\label{T-Q}
\ee
its functional form is illustrated in Fig.\ \ref{R-N}. We note that with
increasing charge, the Coulomb repulsion weakens the gravitational field
at the event horizon and hence decreases the temperature of the corresponding
quantum excitations. As $Q^2 \to GM^2$, the gravitational force is fully
compensated and the black hole is gone. 

\begin{figure}[h]
\centerline{\psfig{file=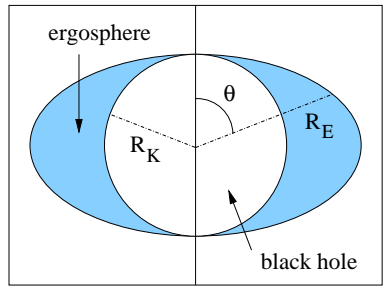,width=5cm}} 
\caption{Geometry of a rotating black hole}
\label{ergo}
\end{figure}

In a similar way, the effect of the angular momentum of a rotating black
hole can be incorporated. It is now the centripetal force which counteracts
the gravitational attraction and hence reduces its strength. The resulting  
Kerr metric must take into account that in this case the rotational symmetry 
is reduced to an axial symmetry, and with $\theta$ denoting the angle relative 
to the polar axis $\theta=0$, it is (at fixed longitude) given by
\be
ds^2 = \left( 1 - {2GMR \over R^2 + a^2 \cos^2 \theta}\right) dt^2 
- {R^2 + a^2 \cos^2 \theta \over R^2 - 2GMR + a^2}~\!dr^2 
- (R^2 + a^2 \cos^2 \theta)~\!d\theta^2.
\label{Kerr}
\ee
The angular momentum of the black hole is here specified by the parameter
$a=J/M$; for $a=0$, we again recover the Schwarzschild case. The general
situation is now somewhat more complex, since eq.\ (\ref{Kerr})
leads to two different divergence points. The solution
\be
R_K = GM~\!(1 + \sqrt{1 - a^2/(GM)^2}~\!)
\label{RK}
\ee 
defines the actual event horizon, corresponding to absolute confinement.
But the resulting black hole is now embedded in a larger ellipsoid
\be
R_E = GM~\!(1 + \sqrt{1 - [a^2/(GM)^2]\cos^2 \theta}),
\label{RE}
\ee 
as illustrated in Fig.\ \ref{ergo}. The two surfaces touch at the poles, and 
the region between them is denoted as the ergosphere. Unlike the black hole
proper, communication between the ergosphere and the outside world is
possible. Any object in the ergosphere will, however, suffer from the
rotational drag of the rotating black hole and thereby gain momentum. 
We shall return to this shortly; first, however, we note that the 
temperature of the Hawking radiation from a rotating black hole becomes
\be
T_{BH}(M,J) = T_{BH}(M,0)~\!\left\{{2 \sqrt{1 - a^2/(GM)^2} 
\over (1 + \sqrt{1- a^2/(GM)^2} }\right\}. 
\label{T-J}
\ee
For a non-rotating black hole, with $a=0$, this also reduces to the 
Hawking temperature for the Schwarzschild case.  

To illustrate the effect of the ergosphere, imagine radiation from a  
Schwarzschild black hole emitted radially outward from the event horizon. 
In the case of a Kerr black hole, such an emission is possible only along 
the polar axis; for all other values of $\theta$, the momentum of the 
emitted radiation (even light) will increase due to the rotational drag 
in the ergosphere. This effect ceases only once the radiation leaves
the ergosphere. Since the amount of drag depends on $\theta$, the momentum 
of the radiation emitted from a rotating black hole, as measured at large 
distances, will depend on the latitude at which it is emitted and increase 
from pole to equator. 

Finally, for completeness, we note that for black holes with both spin and 
charge (denoted as Kerr-Newman), the event horizon is given by
\be
R_{KN} = GM~\!(1 + \sqrt{1 - [Q^2/GM^2] - [a^2/(GM)^2]}),
\label{KN}
\ee 
and the radiation temperature becomes
\cite{Ruf,I-U-W}   
\be
T_{BH}(M,Q,J) = T_{BH}(M,0,0)~\left\{{4 \sqrt{1 - (GQ^2 + a^2)/(GM)^2} 
\over (1 + \sqrt{1- (GQ^2 +a^2)/(GM)^2}~\!)^{~\!2} + a^2/(GM)^2}\right\}. 
\label{T-QJ}
\ee
The decrease of $T_{BH}$ for $Q\not= 0, J\not= 0$ expresses the fact that 
both the Coulomb repulsion and the rotational force counteract the 
gravitational attraction, and if they win, the black hole is dissolved. 

\section{Vacuum Pressure and Baryon Density}

We now want to consider the extension of charged black hole physics to
colour confinement in the case of systems with a net baryon number.
In eq.\ (\ref{T-Q}) we had seen that the reduction of the gravitational
attraction by Coulomb repulsion in a charged black hole in turn reduces
the temperature of Hawking radiation. The crucial quantity here is the
ratio $Q^2/GM^2$ of the overall Coulomb energy, $Q^2/R$, to the overall 
gravitational energy, $GM^2/R$. Equivalently, $Q^2/GM^2 = P_Q/P_G$
measures the ratio of inward gravitational pressure $P_G$ at the event 
horizon to the repulsive outward Coulomb pressure $P_Q$.  

\begin{figure}[h]
\centerline{\psfig{file=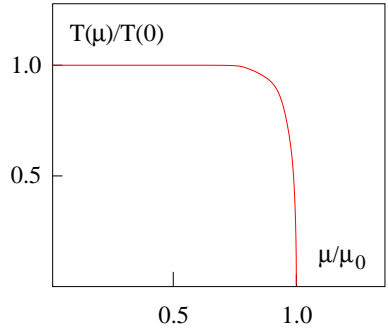,width=5cm}} 
\caption{Hadronization temperature as function of the baryochemical potential}
\label{T-mu}
\end{figure}

In QCD, we have a ``white'' hole containing coloured quarks, confined 
by chromodynamic forces or, equivalently, by the pressure $B$ of the
physical vacuum. If the system has a non-vanishing overall baryon number,
there will be a Fermi repulsion between the corresponding quarks, and this 
repulsion will provide a pressure $P(\mu)$ acting against $B$, with 
$\mu$ denoting the corresponding quark baryochemical potential. We thus
expect a similar reduction of the hadronization temperature as function
of $\mu$. To quantify this, we have to obtain the reduction of the 
chromodynamic force field, such as the string tension $\sigma$, due to
baryonic repulsion. As a first estimate, we start from the simplest picture 
of colour confinement and consider is an ideal gas of massless quarks and 
gluons, held together by the vacuum pressure $B$. At fixed temperature $T$ 
and quark baryochemical potential $\mu$, the overall pressure of such a 
system is given by
\be
P = \left({\pi^2\over 90} \left[d_b + {7 \over 4}d_f \right]\right) T^4 + 
\left({d_f \over 12}\right) \mu^2 T^2
+ \left({d_f \over 24 \pi^2}\right) \mu^4 - B
\label{ideal-qgp}
\ee
where $d_b$ and $d_f$ denote the degrees of freedom of gluons and 
quarks, respectively; for colour $SU(3)$ and two quark flavours,
$d_b=16$ and $d_f=12$. When the inward vacuum pressure just balances
the combined outward kinetic motion of quarks and gluons and the fermi 
repulsion of the quarks, we have $P=0$. For $\mu=0$, this gives
\be
T_0 = \left({90 \over 37 \pi^2}\right)^{1/4}\!B^{1/4} \simeq 0.70~B^{1/4}.
\label{T0}
\ee
On the other hand, for $T=0$, when the Fermi repulsion of the quarks alone 
balances the vacuum pressure, $\mu$ becomes
\be
\mu_0 = (2\pi^2 B)^{1/4} = \pi \left({37 \over 45}\right)^{1/4}\!T_0 
\simeq 3~T_0.
\label{mu0}
\ee 
With a hadronization temperature $T_0 \simeq 0.175$ GeV, this leads to
a baryochemical potential $\mu_b=3\mu \simeq 1.6$ GeV. In the intermediate
region, where both $T$ and $\mu$ are finite, we want to compare the effect
of the Fermi repulsion to the vacuum pressure through the Hawking-Unruh
form, i.e., we replace $Q^2/M^2$ in eq.\ (\ref{T-Q}) by $\mu/\mu_0$, giving 
\be
T(\mu)/T_0 = {\sqrt{1-(\mu/\mu_0)^4} \over (1+\sqrt{1-(\mu/\mu_0)^4})^2}.
\label{Tmu}
\ee
The resulting behaviour of $T(\mu)$ is shown in Fig.\ \ref{T-mu}.

Instead of using the hadronization temperature as scale, the vacuum 
pressure can be determined in terms of the gluon condensate 
$\langle G_{\mu \nu}^2 \rangle$ \cite{svz,shuryak},
\be
B= {33 - 2N_f \over 384 \pi^2} \langle G_{\mu \nu}^2 \rangle.
\label{bag}
\ee
Numerical studies \cite{svz} lead for two flavours to $B^{1/4} \simeq 
0.21 - 0.25$ GeV; from this we get both $T_0 \simeq 0.150 - 0.175$ GeV  
and $\mu_0 \simeq 1.3 - 1.6$ GeV. 

Clearly this approach is overly simplistic, since it reduces the effect
of the additional quarks to only their Fermi repulsion. A more general way 
of addressing the problem would be to introduce an effective $\mu$-dependence 
of the string tension. The presence of further quarks will lead to a
screening-like reduction of the force between a given $\Q$ pair, and hence
a screened string tension, with a $\mu$-dependent screening mass as 
obtained in finite density lattice studies, might provide a more realistic
approach to the $\mu$ dependence of the Hawking-Unruh hadronization 
temperature.

\section{Angular Momentum and Non-Central Collisions}

\begin{figure}[h]
\centerline{\psfig{file=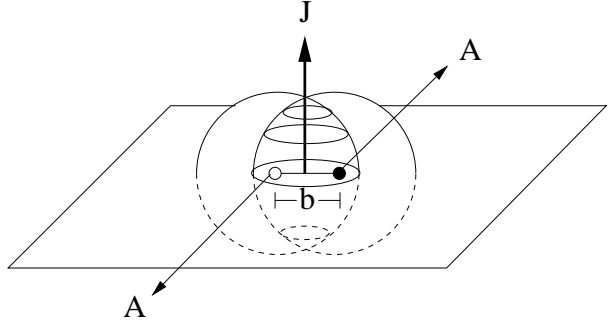,width=8cm}} 
\caption{Rotating interaction region in non-central $AA$ collision} 
\label{spin}
\end{figure}

The dependence of Hawking radiation on the angular momentum of the emitting
system introduces a particularly interesting aspect for the ``white hole 
evaporation'' we have been considering.  Consider a nucleus-nucleus collision
at non-zero impact parameter $b$. If the interaction is of collective 
nature, the resulting system can have some angular momentum orthogonal to
the reaction plane (see Fig.\ \ref{spin}). For central collisions, this
will not be the case, nor for extremely peripheral ones, where one expects
essentially just individual nucleon-nucleon collisions without collective
effects. 

If it possible to consider a kinematic region in which the interacting
system does have an overall spin, then the resulting Hawking radiation
temperature should be correspondingly reduced, as seen in eq.\ (\ref{T-J}).
The effect is not so easily quantified, but simply a reduction of the
hadronization temperature for non-central collisions would quite indicative.
Such a reduction could appear only in the temperature determined by the
relative abundances, since, as we shall see shortly, the transverse 
momentum spectra should show modifications due to the role of the 
ergosphere. The hadronization temperature determined from a resonance
gas study could thus start for central collisions with the value 
(\ref{T-H}) or, at finite $\mu$ with the corresponding $T_H(\mu)$, then 
decrease with the onset of white hole rotation for non-central collisions, 
and finally increase again as collective nuclear effects go away and we 
recover elementary nucleon-nucleon collisions. 

\begin{figure}[h]
\vskip-4.95cm
\hfill{\psfig{file=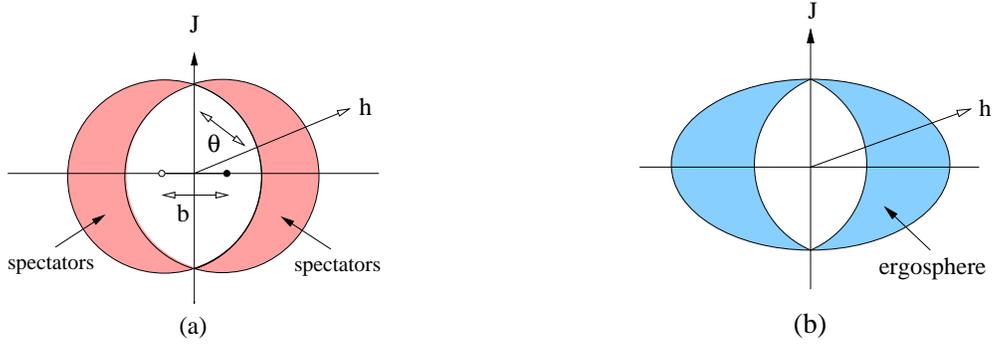,width=4.7cm}}~~~~~~~~~~
\bigskip
\caption{Transverse plane view of a non-central $AA$ collision} 
\label{ellip}
\end{figure}

We next turn to the momentum spectrum of the Hawking radiation emitted
from a rotating white hole. As discussed in section 4, such radiation
will exhibit an azimuthal asymmetry due to the presence of the ergosphere,
which by its rotation will affect the momentum spectrum of any passing
object. At the event horizon, the momentum of all radiation is determined 
by the corresponding Hawking temperature (\ref{T-J}); but the passage of 
the ergosphere adds rotational motion to the emerging radiation and
hence increases its momentum. As a result, only radiation emitted 
directly along the polar axis will have momenta as specified by the
Hawking temperature; with increasing latitude $\theta$ (see Fig.\ 
\ref{ellip})a, the rotation will increase the radiation momentum up to 
a maximum value in the equatorial plane. 

\begin{figure}[h]
~~~~~~~~~~{\psfig{file=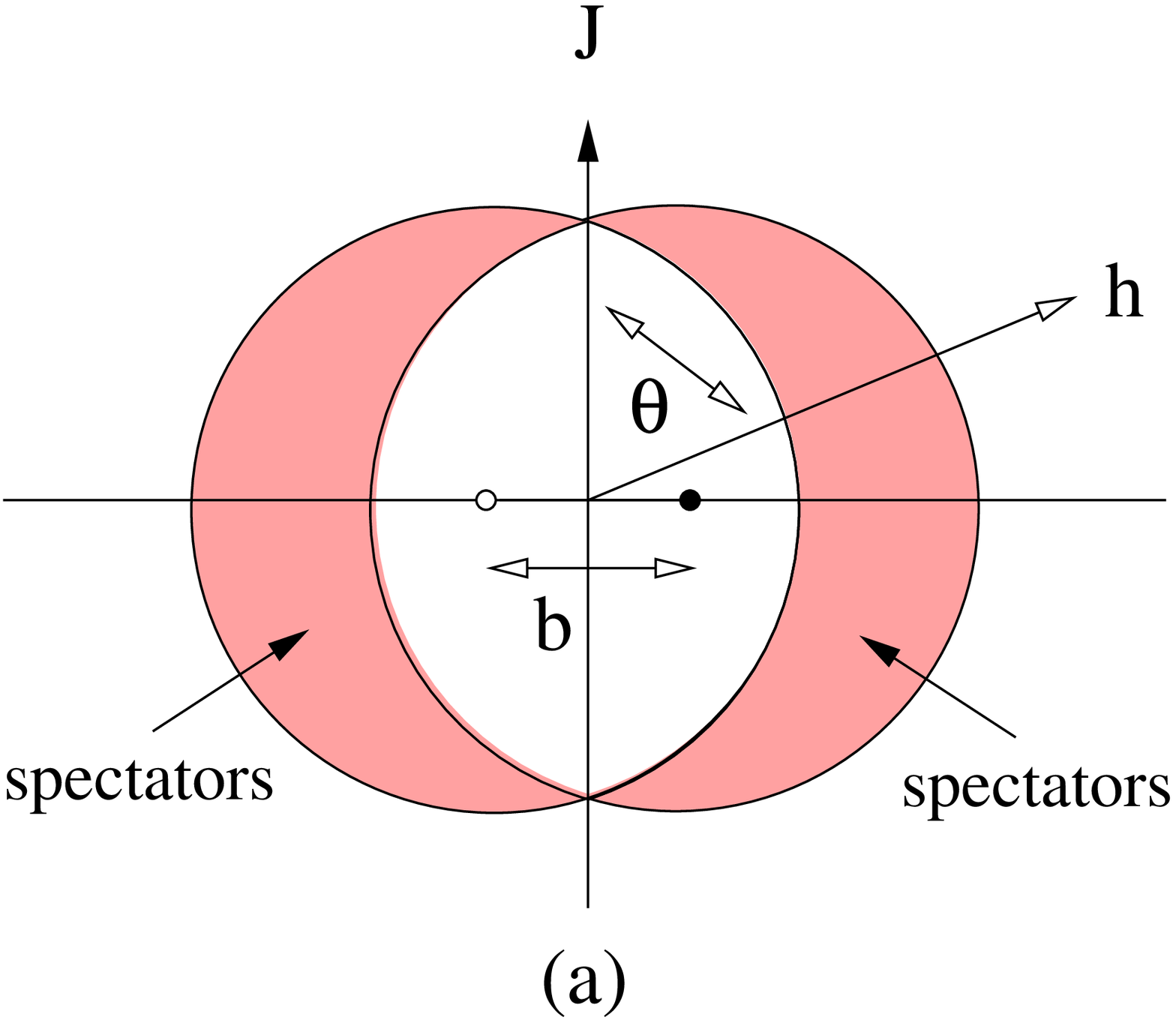,width=5cm}} 
\end{figure}

Hawking radiation from a rotating source thus leads for nuclear collisions 
quite naturally to what in hydrodynamic studies is denoted as elliptic flow. 
It is interesting to note that both scenarios involve collective effects:
while in hydrodynamics, it is assumed that non-central collisions lead to
an azimuthally anisotropic pressure gradient, we have here assumed that
such collisions lead to an overall angular momentum of the emitting 
system.

\section{Stochastic vs.\ Kinetic Thermalization}

In statistical mechanics, a basic topic is the evolution of a system of 
many degrees of freedom from non-equilibrium to equilibrium. Starting
from a non-equilibrium initial state of low entropy, the system is 
assumed to evolve as a function of time through collisions to a 
time-independent equlibrium state of maximum entropy. In other words,
the system loses the information about its initial state through a
sequence of collisions and thus becomes thermalized. In this sense,
thermalization in heavy ion collisions was studied as the transition from
an initial state of two colliding beams of ``parallel'' partons to a final 
state in which these partons have locally isostropic distributions.
This ``kinetic'' thermalization requires a sufficient density of constituents, 
sufficiently large interaction cross sections, and a certain amount
of time.     

From such a point of view, the observation of thermal hadron production 
in high energy collisions in general, in $e^+e^-$ and $pp$ interactions
as well as in heavy ion collisions, is a puzzle: how could these systems 
ever ``have reached'' thermalization? Already Hagedorn had therefore 
concluded that the emitted hadrons were ``born in equilibrium'' 
\cite{Stock}.

Hawking radiation provides a stochastic rather than kinetic approach to
equilibrium. The barrier to information transfer provided by the event
horizon requires that the resulting radiation states excited from the
vacuum are distributed according to maximum entropy, with a temperature 
determined by the strength of the ``confining'' field. The ensemble of
all produced hadrons, averaged over all events, then leads to the same
equilibrium distribution as obtained in hadronic matter by kinetic
equilibration. In the case of a very high energy collision with a high 
average multiplicity already one event can provide such equilibrium;
because of the interruption of information transfer at each of the
successive quantum colour horizons, there is no phase relation between
two successive production steps in a given event. The destruction of
memory, which in kinetic equilibration is achieved through sufficiently 
many successive collisions, is here automatically provided by the
tunnelling process. 

So the thermal hadronic final state in high energy collisions is not
reached through a kinetic process; it is rather provided by successively 
throwing dice.

\section{Conclusions}

We have shown that quantum tunnelling through the colour confinement
horizon leads to thermal hadron production in the form of Hawking
radiation. In particular, this implies:
\begin{itemize}
\item{The radiation temperature $T_H$ is determined by the transverse 
extension of the colour flux tube, giving
\be
T_H = \sqrt{\sigma \over 2 \pi},
\ee
in terms of the string tension $\sigma$.}  
\item{The multiplicity $\nu(s)$ of the produced hadrons is determined 
by the increase of the flux tube thickness with string length, leading to  
\be
\nu(s) \simeq \ln \sqrt s,
\ee
where $\sqrt s$ denotes the collision energy.}

\item{The temperature of Hawking radiation can in general depend on the 
charge and the angular momentum of the emitting system. The former provides 
the dependence of the hadronization temperature on baryon density. Given 
sufficiently much collective collision behaviour, the latter could cause 
elliptic flow and a centrality dependence of
$T_H$ in $AA$ collisions.} 
\item{In statistical QCD, thermal equilibrium is reached kinetically
from an initial non-equilibrium state, with memory destruction through
successive interactions of the constituents. In high energy collisions, 
tunnelling prohibits information transfer and hence leads to stochastic 
production, so that we have a thermal distribution from the outset.} 
\end{itemize}

\end{document}